\documentclass[sigconf]{acmart}
\usepackage{tikz}
\newcommand*\circled[1]{\tikz[baseline=(char.base)]{
            \node[shape=circle,draw,inner sep=1pt] (char) {#1};}}

\AtBeginDocument{%
  \providecommand\BibTeX{{%
    \normalfont B\kern-0.5em{\scshape i\kern-0.25em b}\kern-0.8em\TeX}}}

\copyrightyear{2022} 
\acmYear{2022} 
\setcopyright{acmcopyright}\acmConference[ICSE-NIER'22]{New Ideas and Emerging Results }{May 21--29, 2022}{Pittsburgh, PA, USA}
\acmBooktitle{New Ideas and Emerging Results (ICSE-NIER'22), May 21--29, 2022, Pittsburgh, PA, USA}
\acmPrice{15.00}
\acmDOI{10.1145/3510455.3512788}
\acmISBN{978-1-4503-9224-2/22/05}

\begin{document}

\title{Towards a Reference Software Architecture for Human-AI Teaming in Smart Manufacturing}

\author{Philipp Haindl}
\email{philipp.haindl@scch.at}
\orcid{0000-0001-6075-5286}
\affiliation{%
  \institution{Software Competence Center Hagenberg}
  \city{Hagenberg}
  \country{Austria}
}

\author{Georg Buchgeher}
\email{georg.buchgeher@scch.at}
\orcid{0000-0002-8565-6257}
\affiliation{%
  \institution{Software Competence Center Hagenberg}
  \city{Hagenberg}
  \country{Austria}
}

\author{Maqbool Khan}
\email{maqbool.khan@fecid.paf-iast.edu.pk}
\orcid{0000-0001-7656-0184}
\affiliation{%
  \institution{Pak-Austria Fachhochschule - Institute of Applied Sciences and Technology}
  \city{Mang, Haripur}
  \country{Pakistan}
}

\author{Bernhard Moser}
\email{bernhard.moser@scch.at}
\orcid{0000-0003-1859-046X}
\affiliation{%
  \institution{Software Competence Center Hagenberg}
  \city{Hagenberg}
  \country{Austria}
}

\renewcommand{\shortauthors}{}
\begin{abstract}
With the proliferation of AI-enabled software systems in smart manufacturing, the role of such systems moves away from a reactive to a proactive role that provides context-specific support to manufacturing operators. In the frame of the EU funded \textit{Teaming.AI} project, we identified the monitoring of teaming aspects in human-AI collaboration, the runtime monitoring and validation of ethical policies, and the support for experimentation with data and machine learning algorithms as the most relevant challenges for human-AI teaming in smart manufacturing. Based on these challenges, we developed a reference software architecture based on knowledge graphs, tracking and scene analysis, and components for relational machine learning with a particular focus on its scalability. Our approach uses knowledge graphs to capture product- and process specific knowledge in the manufacturing process and to utilize it for relational machine learning. This allows for context-specific recommendations for actions in the manufacturing process for the optimization of product quality  and the prevention of physical harm. The empirical validation of this software architecture will be conducted in cooperation with three large-scale companies in the automotive, energy systems, and precision machining domain. In this paper we discuss the identified challenges for such a reference software architecture, present its preliminary status, and sketch our further research vision in this project. 
\end{abstract}


\maketitle

\section{Introduction}
Applications of AI in smart manufacturing are manifold, ranging from improving maintenance times of machinery, the detection of failures in the product or the machinery to the prevention of harm to manufacturing operators. In general, complex processes that are worked on collaboratively are characterized by a sequence of reactive and proactive elements, which each actor alternatingly supporting the other. AI-enabled systems in smart manufacturing are capable of self-sensing, self-adaptation, self-organization, and self-decision \cite{qu_smart_2019,phuyal_challenges_2020}, allowing them to respond to physical changes in the production environment in various ways - by stopping machines, adapting production tasks, or suggesting the change of production parameters. However, effective teaming between manufacturing operators and AI-enabled manufacturing systems requires mutual trust into the other's capabilities, primarily resulting from self-sensing and self-adaption. With regards to collaborative AI systems, this demands a high degree of situational awareness for each other actor's needs, knowledge of the production process, and its adjustable parameters. 

From a higher perspective, two main challenges related to teaming AI and manufacturing operators in smart manufacturing can be identified: The first challenge relates to the required scalability of the architecture when processing data in near-realtime, particulary in combination with relational machine learning, i.e., the statistical analysis of relational, or graph-structured, data \cite{nickel_review_2016}. The second challenge relates to examining a suitable framework to explicate the knowledge for effective teaming in the manufacturing process. \textit{Shared mental models} capture the common ground knowledge in the collaboration between humans with robots \cite{scheutz_framework_2017,gervits_shared_2018,jonker_shared_2011}. We use knowledge graphs and ontologies to formalize these shared mental models of the manufacturing process and the semantics of trust factors for human-AI teaming in an operational manner. 

In this paper we first present the initial reference software architecture as one important contribution of the \textit{Teaming.AI}\footnote{https://www.teamingai-project.eu} project. The remainder of the paper is structured as follows: First we sketch the envisioned approach based on a use case from one of the industrial partners in Section \ref{illustrativeexample}, before we elaborate the identified challenges for such a reference architecture in Section \ref{challenges}. Afterwards, we present the current status of this software architecture in Section \ref{results}. Finally, we sketch our next research activities and conclude the paper in Section \ref{conclusion}.

\section{Envisioned Approach}
\label{illustrativeexample}
One of our industrial partners specializes in high-precision machining of large-sized parts by milling or grinding based on cast materials or machine-welded structures. Manufacturing operators must manually clamp large and heavy parts into high-precision manufacturing machines to perform grinding or milling operations. This process takes up a large portion of the total cycle time of a work order and workers are exposed to occupational hazards. A reliable full automation by an AI-assisted robotic system is beyond the capabilities of current AI-based technologies, due to large part variety and the complexity of the associated handling processes. 

By compensating for the limited visual perception of both actors in the manufacturing process - the manufacturing operator and the AI system - we seek to maximize awareness of hazardous situations and conditions that negatively impact physical or mental ergonomics. Based on human feedback, the AI system learns to predict which action sequences are ergonomically favorable. However, since the AI-augmented manufacturing system cannot be sure that it has the complete information about the situation due to occlusions and unseen areas, it expresses its confidence to the manufacturing operator by, for example, displaying safety warnings. The system fuses observational data from a visual tracking system with prior knowledge about the scene and process with in-process feedback from the manufacturing operator. Obviously, for such a system to gain the trust from the manufacturing operator, not only the correctness, and context-dependent suitability are critical success factors, but also the timely provision of recommended actions or the display of safety warnings. Of course, such a system must also comply with ethical policies and standards, which the manufacturing operator also tacitly assumes. 

The general approach envisioned in our research project uses these data for machine learning and the development of a context-aware AI-augmented manufacturing system. Thereby, prior knowledge about the scene, geometry of the parts, and compositional workflow patterns for the handling and processing steps are encoded in a knowledge graph, along with safety guidelines. The knowledge graph is updated at runtime with current contextual information. An integrated self-diagnostic component simultaneously assesses the system's level of consistency and completeness and expresses its epistemic confidence. In case of ambiguity, the manufacturing operator is asked for feedback. 

From a functional perspective, the teaming of manufacturing operators and AI in an industrial manufacturing process raises specific challenges related to relational machine learning, processing IIoT streaming data, the modeling and monitoring of trust (factors), and the continuous evaluation of human feedback and compliance with ethical policies. From a non-functional perspective, scalability and timing requirements of the components play a special role for meeting the functional requirements in such a system.

\section{Research Challenges}
\label{challenges}
In multiple workshops over the course of six months our research consortium identified five core challenges relevant for a reference architecture for human-AI teaming in smart manufacturing. These workshops involved researchers from software engineering (3), knowledge engineering (3), machine learning (4), computer vision (2), vision systems (2), and human factors (1) and domain experts from the companies, such as production and quality managers. 
\\
\textbf{\circled{1} Monitoring of Teaming Aspects in Human-AI Interaction.} Salsas et al. in \cite{salas_is_2005} identified five core components that are necessary for effective teamwork. In addition, shared mental models \cite{jonker_shared_2011}, mutual trust, and closed-loop communication serve as coordination mechanism between the actors. We use these concepts as foundation for modeling the manufacturing process as alternating activities between the manufacturing operator and the AI-augmented manufacturing system. Further, we rely on the \textit{4S Interdependence Framework} to define the neccessary state, structural organization, skills, and strategy for each actor and activitiy in the smart manufacturing process. The monitoring of these aspects, e.g., through mining event logs or scene detection through a tracking and scene analysis system, is the main challenge in this context.
\newline 
\textbf{\circled{2} Scalability for Near-Realtime IIoT Data Processing. }
In smart manufacturing, streaming data acquired from IIoT sensors continuously need to be processed so that they can be used for machine learning. Thus, the scalability of a reference architecture \cite{khalil_deep_2021,santos_systems_2019} remains a key challenge in this context as delayed or incorrect recommendations of the AI systems to manufacturing operators may delay the workflow or, even worse, can also endanger their safety. In this context we research on appropriate consistency requirements in the face of data sharding between different replicas to better cope with typical data volumes in this context.
\newline
\textbf{\circled{3} Runtime Monitoring and Validation of Ethical Policies} 
Formalizing ethical and standards in an operational manner and validating their fulfillment by an AI system at runtime remains a key challenge \cite{vakkuri_ethically_2019,vakkuri_eccola_2021,hagendorff_ethics_2020,johnson_towards_2021} in this context. 
The continuous validation of fulfillment of these policies and standards is exacerbated by the fact that they can only be evaluated at runtime. In our research we aim on developing a formalism based on ontology-reasoning to express these policies and standards in an operational manner.
\newline
\textbf{\circled{4} Relational Machine Learning for Knowledge Graphs.} 
The first research challenge relates to examining different approaches for calculating graph embeddings and their respective computational costs \cite{nickel_review_2016,trouillon_knowledge_2017}. Second, for the development of the framework abstraction layer, the required functionalities for orchestrated machine learning need to be identified, which is the main challenge in this context. Subsequently, generic interfaces covering these common functionalities can be derived thereof. 
\newline
\textbf{\circled{5} Experimentation of Data and Algorithms.} 
The automatization of experiments requires an operational definition of the experiments' evaluation criteria, the parameterization of required data pipelines and machine learning algorithms, and the configuration of their deployment and execution. Thus, the research challenge comprises the development of domain-specific languages to describe and evaluate these experiments based on quality factors of AI-based systems \cite{siebert_towards_2020} in an operational manner. Further it comprises the development of respective tooling for automatizing the evaluation of the experiments \cite{nguyen-duc_continuous_2020,bosch_engineering_2020} that also integrates qualitative expert feedback from the human-in-the-loop (HITL).

\section{Preliminary Status of the Reference Software Architecture}
\label{results}
\begin{figure*}[h!]
  \centering
  \includegraphics[scale=0.38]{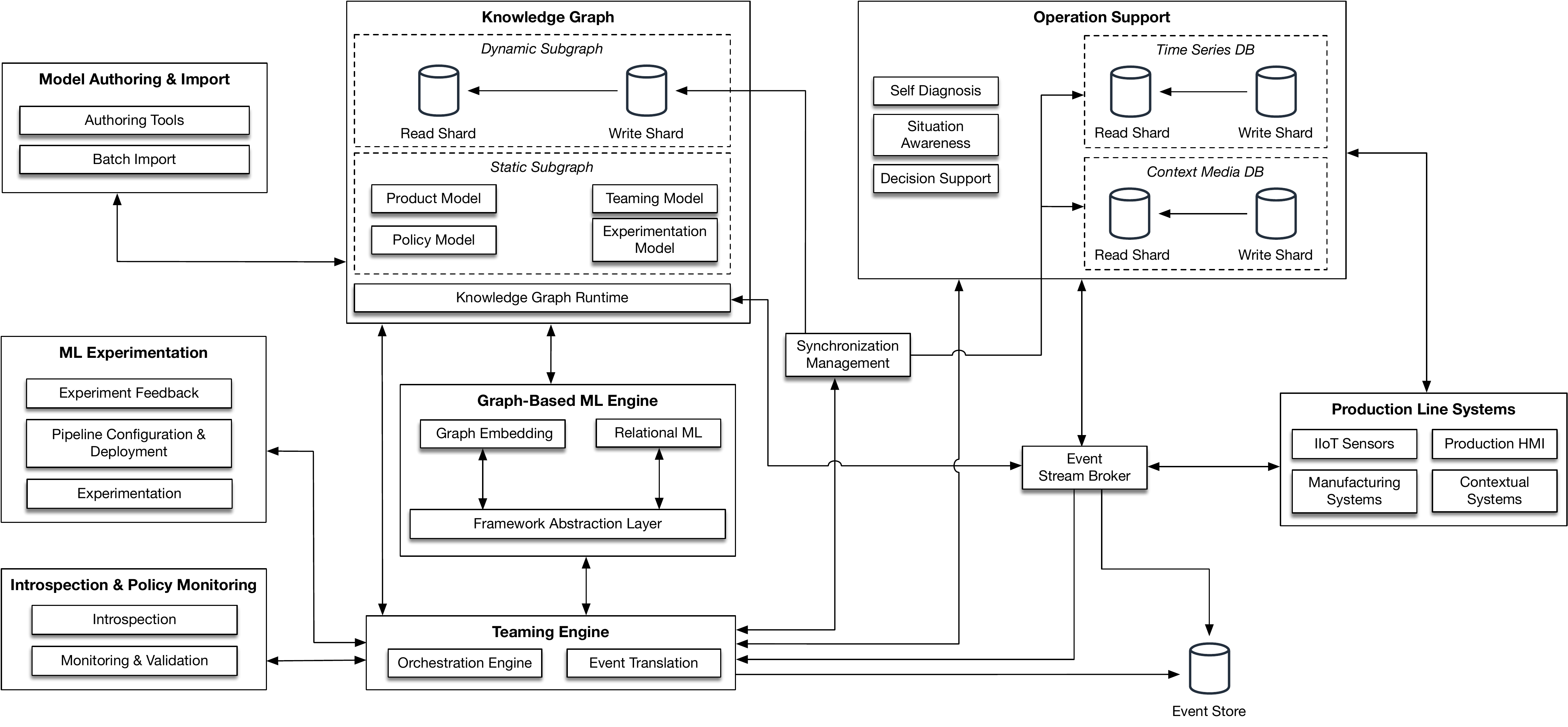}
  \caption{Reference Software Architecture For Human-AI Teaming in Smart Manufacturing.\label{metaarchitecture}}
  \Description{Reference Software Architecture For Human-AI Teaming in Smart Manufacturing.}
\end{figure*}
Based on the aforementioned challenges our research consortium developed a reference software architecture that serves as a blueprint for our subsequent research activities and validations. Though this architecture merges different viewpoints from researchers with software engineering and machine learning backgrounds, we expect subtle changes with progress of the research project. Thus, this description captures its preliminary status. 

Figure \ref{metaarchitecture} shows the different components of this reference architecture. To account for the different latency requirements of the components to process the data in a streaming-like manner, we followed the \textit{Lambda architecture} pattern as described by Warren and Marz \cite{warren_big_2015}. This architectural pattern groups the components based on their latency requirements into three layers. The \textit{batch layer} (model authoring) ingests and stores large amounts of data, the \textit{speed layer} (knowledge graph, graph-based ML and teaming engine, production line systems) processes updates to the data in low-latency, and the \textit{serving layer} (operation support, ML experimentation, introspection \& policy monitoring) provides precalculated results also in a low-latency fashion \cite{preuveneers_samurai_2016}. To separate read and write operations and therewith be able to balance the processing of large data volumes, all data stores used in the architecture (i.e., dynamic knowledge graph, time series, and media data) are replicated as \textit{read} and \textit{write shards}. The synchronization between these replicas is performed autonomously by the \textit{synchronization management} component. 

\subsection{Model Authoring \& Import}
Product- and process specific models for the manufacturing process are imported into the knowledge graph by means of batches. In this regard, \textit{importing} might also comprise to convert these domain-specific models into a graph representation such as RDF (Resource Description Framework). This component also embodies the functionality needed for authoring and versioning the models.
\subsection{ML Experimentation}
During the continuous improvement and testing of AI-based systems, huge amounts of data are generated that can be used for targeted experiments with these systems \cite{arpteg_software_2018,amershi_software_2019}. Particularly, the indeterministic nature of data-driven algorithms and their entanglement with the used training data underpins the need for continuous experimentation and validation of AI components. 

\subsection{Introspection \& Policy Monitoring}
Explainability of AI (XAI) \cite{rudin_stop_2019}, ethics-based auditability \cite{mokander_ethics_2021}, and the alignment with human rights and autonomy \cite{ieee_ieee_2019} are common examples for ethical requirements towards AI systems. In recent years, several policies and guidelines have been presented by companies, e.g., Google \cite{pichai_ai_2018}, governmental bodies, e.g., the EU \cite{hleg_ethics_2019}, and standardization organizations, e.g., from IEEE \cite{ieee_ieee_2019}. The compliance with these standards and policies needs to be assured throughout the operation of any AI-enabled software system. To this end, this component encapsulates \textit{introspection} capabilities, i.e., the self-directed evaluation of the AI system by itself, and additional monitoring and validation facilities related to ethical standards and policies. While introspection is triggered ad-hoc, e.g., upon the detection of suspicious interaction patterns, monitoring and validation is performed continuously at runtime. To also take into account historical data when evaluating the compliance with ethical policies, all processed events are stored in the \textit{event store}. 

\subsection{Knowledge Graph}
Due to the emerging use of knowledge graphs to represent ontologies and semantic data, they also become increasingly important to formalize expert knowledge, process, and simulation data in manufacturing \cite{listl_knowledge_2020}. Based on the frequency of updates to the data, the reference architecture differentiates between a dynamic and a static knowledge subgraph. Models related to the product, the manufacturing process, and to experimentation are static in the sense that they do not need to be continuously updated at runtime. These data are embodied in the \textit{static subgraph}. The \textit{dynamic subgraph} on the other hand covers operational data accruing in the manufacturing process itself \cite{ringsquandl_filling_2018,ringsquandl_event-driven_2017}, augmented by added facts from relational machine learning to optimize the interplay between human and AI. This interplay can, e.g., be optimized by proactively giving the operator recommendations for the next manufacturing step or suggesting parameter changes for the machinery to respond to observed product quality deviations, detected by the AI systems. 

\subsection{Graph-based ML Engine}
This component provides \textit{relational machine learning} capabilities that focus on the application of machine learning methods onto relational or graph-structured data such as knowledge graphs. An essential prerequisite for this is the calculation of \textit{graph embeddings}, i.e., the transfer of a graph representation into a vector space. to orchestrate complex machine learning tasks (e.g., via frameworks such as \textit{Kubeflow}, \textit{TensorFlow Extended}, or \textit{MLFlow}) it provides a \textit{framework abstraction layer} to make these frameworks accessible through a generic software interface.

\subsection{Teaming Engine}
The \textit{orchestration engine} controls the reliable execution of the teaming process, which is defined in the \textit{teaming model}. This model structures the concrete sequence of activities in the interaction between the AI system and the manufacturing operator. Also, it gives a framework to define which policies are relevant in a specific time period of the manufacturing process and to formalize the ground rules for team interaction, as described by the \textit{4S Interdependence Framework} of Johnson and Vera \cite{johnson_no_2019}. For the orchestration of the teaming process, two types of events are processed by the teaming engine: \textit{Raw events} are extracted from the production line system through IIoT sensors, whereby \textit{teaming events} are exchanged by the different components after processing. As a result, only teaming events contain contextual data. The \textit{orchestration engine} can only work with teaming events, since the contextual data must also be evaluated to decide on the next manufacturing activity. The \textit{event translation} merges several atomic events so that they can be used for decision-making by the orchestration engine.

\subsection{Operation Support}
The current situation on the shopfloor, i.e., the position of parts and movements of the manufacturing operator,  machine settings, and their observed effects on production quality must be taken into account when giving recommendations to the manufacturing operator. Thereby, tracking and scene analysis (TSA) is provided by the \textit{situation awareness} component, based on media data such as video, imagery or audio recordings. The system's \textit{self diagnosis} evaluates the reliability of its recommendations and prematurely alerts upon violations of policies. The interplay of these components, the knowledge graph, and machine learning components for the calculation and selection of recommended actions to the manufacturing operator is controlled by the \textit{decision support} component.

For the storage of near-realtime data acquired from the production line systems, a time series database is provided. Media data from the TSA system are persisted in the \textit{context media database}. 

\subsection{Production Line Systems}
Operational data from production and contextual systems, e.g., auxillary systems or manufacturing process control, are acquired from IIoT sensors or from machine-specific implementations, e.g., based on the OPC Unified Architecture \cite{opc_foundation_what_2008}. An HMI (Human Machine Interface) is provided to interact with the manufacturing operator, to gather context-dependent feedback about the relevance of given recommendations by the AI. Also, it allows to acquire information about unexpectedly taken actions by the manufacturing operator.
 
\section{Future Work and Conclusion}
\label{conclusion}
The presented reference software architecture is framework- and technology agnostic. It shall serve as a blueprint for deriving the software architecture for a particular manufacturing context in a producing company. We use this reference architecture to derive the concrete architecture for the AI-enabled smart manufacturing systems of our three industrial partners in the automotive, energy systems, and precision machining domain. Thus, all three companies have different manufacturing processes in place that result in diverging scalability and interoperability requirements of the software architecture. The validation of the reference architecture will investigate its applicability to different manufacturing contexts, its scalability, and its suitability for validating the compliance of ethical standards during operation. To examine its applicability we will conduct expert interviews with software architects of our industrial partners after having applied the reference software architecture. Likewise, we will conduct interviews with manufacturing operators to assess the suitability of the overall approach and how well it supports the teaming between human and AI in smart manufacturing. Complementary, the quantitative validation will use data acquired from runtime probes to examine the scalability and the consistency among the data shards under heavy load. 

\section{Acknowledgements}
This project has received funding from the European Union’s Horizon 2020 research and innovation
programme under grant agreement No 957402.

\bibliographystyle{ACM-Reference-Format}
\bibliography{bib}

\end{document}